\documentclass{PoS}

\newcommand{\gev}{\,{\rm GeV}}
\usepackage{slashed}

\title{Timelike deeply virtual Compton scattering with a linearly polarized real (or quasi-real) photon beam}

\ShortTitle{Linearly polarized timelike  DVCS }

\author{ A.T. Goritschnig\\
Institute of Physics - Theory Division, University of Graz, Graz, Austria}
\author{B. Pire\\
CPHT, {\'E}cole Polytechnique, CNRS, 91128 Palaiseau, France}
\author{\speaker{Jakub Wagner}\\
        National Center for Nuclear Research (NCBJ), Warsaw, Poland\\
        E-mail: \email{jwagner@fuw.edu.pl}}

\abstract{We calculate timelike  virtual Compton scattering amplitudes  in the generalized Bjorken scaling regime and focus on a new polarization asymmetry in 
the scattering process with a linearly polarized photon beam in the medium energy range, which will be studied intensely at JLab12 experiments. 
We demonstrate that new observables  help us to access the polarized quark and gluon generalized parton distributions $\tilde H(x, \xi, t)$ and $ \tilde E(x, \xi, t)$.}

\FullConference{XXII. International Workshop on Deep-Inelastic Scattering and Related Subjects,\\
		28 April - 2 May 2014\\
		Warsaw, Poland}

\begin{document}

%
%
\maketitle
\thispagestyle{empty}
\renewcommand{\thesection}{\arabic{section}}

\renewcommand{\thesubsection}{\arabic{subsection}}
\noindent
\section{Introduction}
The spin structure of the nucleon remains after decades of experimental efforts an open question of hadronic physics; in particular the contribution of gluons to the nucleon spin stays at a rather unsatisfactory qualitative level, even for integrated parton distributions. A more refined version of this puzzle may be addressed through the  polarized generalized parton distributions (GPDs)  both for the parton helicity conserving ones $\tilde H(x, \xi, t)$ and $ \tilde E(x, \xi, t)$, and the parton helicity flip  ones   $H^T(x, \xi, t),  E^T(x, \xi, t),\tilde H^T(x, \xi, t)$ and $ \tilde E^T(x, \xi, t)$.  Accessing them through various  deep exclusive processes turns out to be more difficult than anticipated. Experimental techniques have recently been developed to yield a linearly polarized, intense photon beam at JLab, using a forward tagger to detect electrons scattered at very small angles at CLAS12 \cite{CLAS12} or using coherent bremsstrahlung technique at the GlueX experiment \cite{Somov:2011zz}. 

We recently demonstrated \cite{GPW} that such a set up will allow to perform new tests of the polarized quark and gluon GPDs  $\tilde H(x, \xi, t)$ and $ \tilde E(x, \xi, t)$, through the study of timelike Compton scattering (TCS) if proper observables are measured.
We  investigated, at leading twist and next to leading $\alpha_S$ order, the photoproduction of lepton pairs off a proton where both a timelike deeply virtual Compton scattering and a Bethe-Heitler (BH) process contribute. When studying their interference one gets direct access to the amplitude of the Compton process. Through it one gets information on the nucleon structure in terms of GPDs. We introduce new observables which are sensitive to different angular dependencies of the outgoing leptons  to project out cross-section contributions associated with particular GPDs. We focus on the medium energy range which will be studied intensely at JLab12 experiments \cite{Nadel-Turonski:2012zba} .

The hard exclusive photoproduction of a lepton pair through TCS\cite{BDP,PSW}
\begin{equation}
\gamma(q_{in}) N(p) \to \gamma^*(q_{out}) N' (p') \to l^-(k)l^+(k')N'(p')\,,
\label{process}
\end{equation}
shares many features with deeply virtual Compton scattering (DVCS)\cite{historyofDVCS}, in the sense that it obeys a factorization property in the generalized Bjorken limit of large $Q^2$ and fixed ratio ${Q^2}/{s}$ with
\begin{equation}
q_{in}^2 =0~~~;~~q_{out}^2 =Q^2 >0 ~~~;~~s=(p+q_{in})^2 \,.
\end{equation} 
These two reactions allow one to access the same knowledge of the nucleon structure encoded in universal GPDs\cite{gpdrev}. Their amplitudes  are related at Born order by a simple complex conjugation, but they significantly differ  \cite{NLOTCS} at next to leading order (NLO) in the strong coupling constant $\alpha_s$. Moreover, the discovery  \cite{Moutarde:2013qs} of the importance of NLO corrections driven by gluonic GPDs allows to get an easy access to the gluonic content of nucleons in spacelike and timelike DVCS reactions, although they do not contribute at the Born order. These two processes are the most promising ones to access the three-dimensional partonic picture of the nucleon \cite{Impact} with the help of the extraction of quark and gluon GPDs.

We consider the process in which a linearly polarized real photon interacts with an unpolarized proton and produces a lepton pair. To specify the variables we use in the description of this process we  introduce two coordinate systems. The first one, similar to the one introduced in \cite{BDP}, is the $\gamma p$ center-of-momentum frame. In the case of a linearly polarized photon we now have a distinguished transverse direction given by the polarization vector, which we choose to point in the $x$-direction. We define 
the scattering angle $ \theta_{c.m.}$  through:
\begin{equation}
\sin \theta_{c.m.} = \frac{|\vec{p}\times \vec{p}~'|}{|\vec{p}||\vec{p}~'|} = \frac{\Delta_T}{|\vec{p}~'|} \sim \mathcal{O}\left(\sqrt{\frac{-t}{s}}\right) \,,
\end{equation}
and the angle $\Phi_h$ between the polarization vector and the hadronic plane as:
\begin{equation}
\sin \Phi_h = \vec{\epsilon}(q_{in})\cdot \vec{n} =  \vec{\epsilon}(q_{in})\cdot \frac{\vec{p}~'\times \vec{p}}{|\vec{p}~'\times \vec{p}|}\,,
\end{equation}
where $\vec{n}$ is the vector normal to the hadronic plane.
Two additional angles  $\theta$ and $\phi$ are defined, as in \cite{BDP}, as the polar and azimuthal angles of $\vec{k}$ in the $l^+l^-$ center-of-momentum frame with the $z$-axis along $-\vec{p}~'$, and the other axes such that $\vec{p}$ lies in the $x$-$z$ plane and has positive $x$-component.

The  full Compton scattering amplitude in its factorized form is conveniently expressed through Compton Form Factors $\mathcal{H}, \mathcal{E},\tilde{\mathcal{H}}, \tilde{\mathcal{E}}$:
\begin{eqnarray}
\mathcal{A}^{\mu\nu}(\eta,t) &=& - e^2 \frac{1}{(p+p')\cdot n_-}\, \bar{u}(p^{\prime}) 
\bigg[\,
g_T^{\mu\nu} \, \Big(
      {\mathcal{H}(\eta,t)} \, \slashed n_- +
      {\mathcal{E}(\eta,t)} \, \frac{i \sigma^{\alpha\rho}n_{-\alpha}\Delta_{\rho}}{2 M}
   \Big) \nonumber \\
& & \phantom{AAAAAAAAA}
   +i\epsilon_T^{\mu\nu}\, \Big(
    {\tilde{\mathcal{H}}(\eta,t)} \, \slashed n_-\gamma_5 +
      {\tilde{\mathcal{E}}(\eta,t)} \, \frac{\Delta^{+\alpha}n_{-\alpha}\gamma_5}{2 M}
    \Big)
\,\bigg] u(p) \,,
\label{eq:amplCFF}
\end{eqnarray}
where :
\begin{eqnarray}
\left\{\mathcal{H}, \mathcal{E}\right\}(\eta,t) &=& + \int_{-1}^1 dx \,
\left(\sum_q T^q(x,\eta)\left\{H^q,E^q\right\}(x,\eta,t)
 + T^g(x,\eta)\left\{H^g,E^g\right\}(x,\eta,t)\right) \nonumber \\
\left\{\tilde \mathcal{H}, \tilde \mathcal{E}\right\}(\eta,t) &=& - \int_{-1}^1 dx \,
\left(\sum_q \tilde T^q(x,\eta)\left\{\tilde H^q,\tilde E^q\right\}(x,\eta,t) 
+\tilde T^g(x,\eta)\left\{\tilde H^g,\tilde E^g\right\}(x,\eta,t)\right).
\label{eq:CFF}
\end{eqnarray}
The renormalized coefficient functions $T^{q,g}, \tilde{T}^{q,g}$ calculated at LO and NLO in $\alpha_S$ can be found in \cite{Moutarde:2013qs}. The scaling skewness variable $\eta$ is defined as:
\begin{eqnarray}
\eta &\equiv& -\frac{(q_{in}-q_{out})\cdot(q_{in}+q_{out})}{(p+p')\cdot(q_{in}+q_{out})} \approx \frac{Q^2}{2s-Q^2} \,.
\end{eqnarray}
\section{Cross sections and observables}

\begin{figure}
\includegraphics[keepaspectratio,width=0.45\textwidth,angle=0]{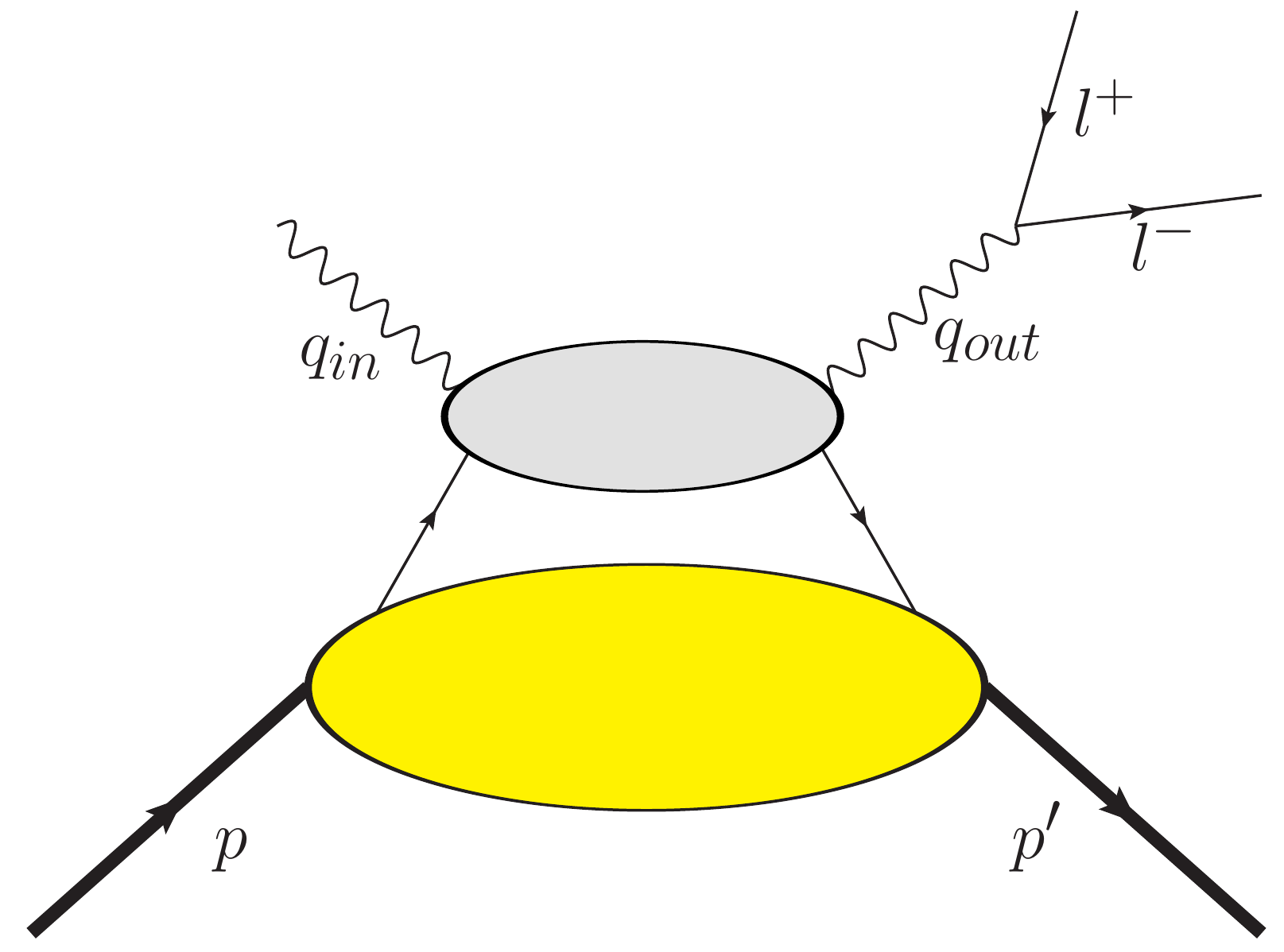} 
\hspace{0.05\textwidth}
\includegraphics[keepaspectratio,width=0.35\textwidth,angle=0]{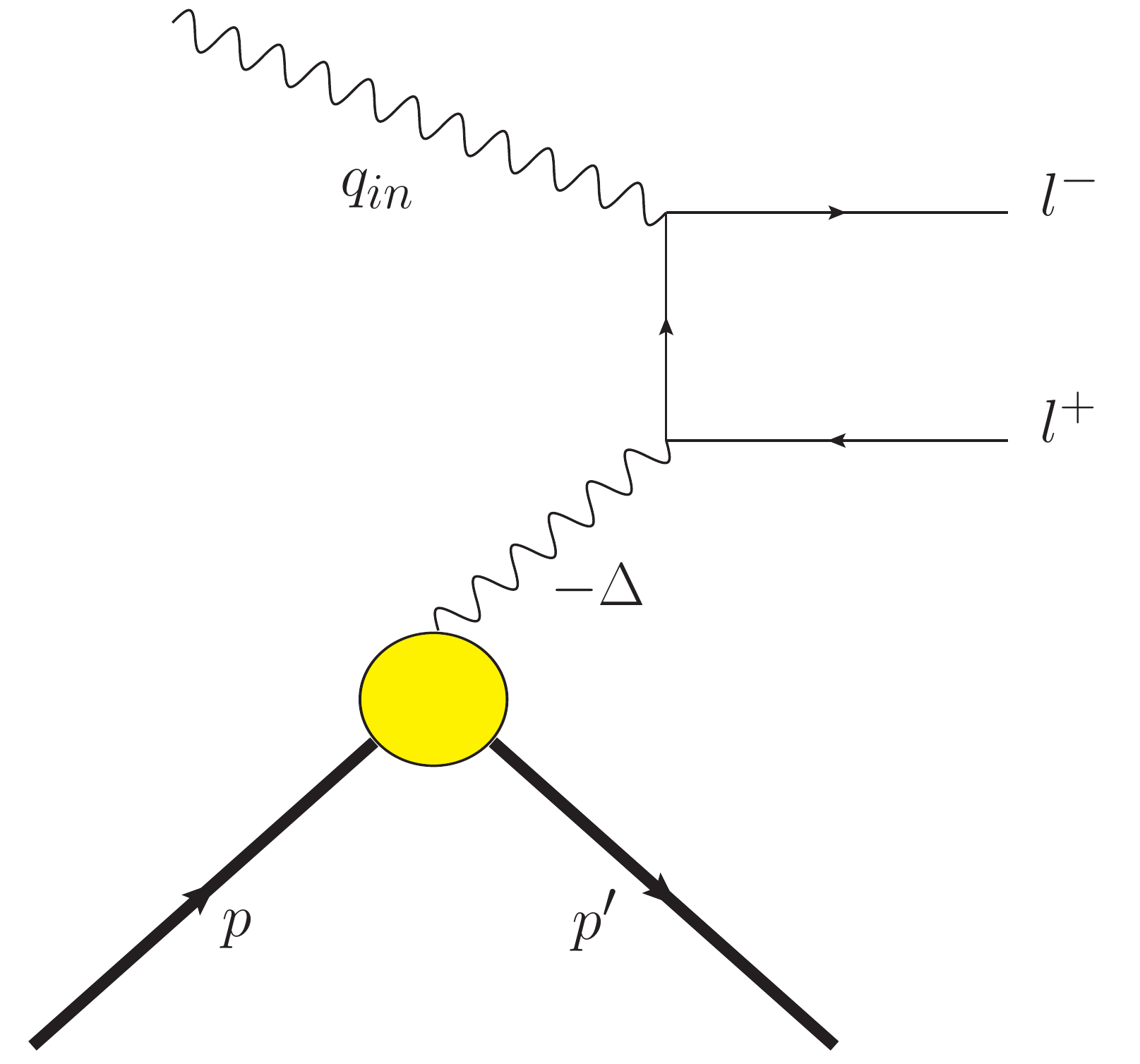} 
\caption{Two mechanisms contributing to the  photoproduction of lepton pairs: timelike Compton scattering (Left) and Bethe-Heitler (Right).}
\label{fig:BH_TCS}
\end{figure}

The physical process where we study TCS is the exclusive photoproduction of a lepton pair. 
But to this process  contribute both TCS and a BH mechanism, where 
 the incoming photon fluctuates into the lepton pair, which then goes on-shell 
via the exchange of a virtual photon with the proton, cf. figure \ref{fig:BH_TCS}. 
It turns out that the BH contribution  dominates over the TCS one in the whole kinematical range under consideration but the quantum mechanical interference of these two amplitudes allows us to extract information on the Compton process itself. 
The interference part can be accessed through the study of the angular distribution of the produced lepton pair.

Up to the leading order accuracy the contribution of the TCS mechanism to the differential $\gamma p \to l^+ l^- p$ photoproduction cross section for fixed polarization of the incoming photon reads (we average over the initial proton and sum over all final particle polarizations):
\begin{eqnarray}
&&\frac{d\sigma^{(TCS)}}{dQ^2 dt d\Omega_{l^+l^-} d \Phi_h} 
 \\
& \,  = \,  & 
\frac{\alpha^3}{16\pi^2s^2}\frac{1}{Q^2}\left( 1 - \sin^2\theta \cos^2(\Phi_h + \phi ) \right) 
\left[ (1-\eta^2) \mid \mathcal{H} \mid^2 - (\eta^2+\frac{t}{4M^2}) \mid \mathcal{E} \mid^2 - 2\eta^2 {Re}( \mathcal{H^*} \mathcal{E}) \right] \nonumber\\
  &+&\frac{\alpha^3}{16\pi^2s^2}\frac{1}{Q^2}\left( 1 - \sin^2\theta \sin^2(\Phi_h + \phi ) \right)  
\left[ (1-\eta^2) \mid\tilde{\mathcal{H}}\mid^2 - \eta^2\frac{t}{4M^2} \mid\tilde{\mathcal{E}}\mid^2 - 2\eta^2{Re}(\tilde{\mathcal{H}}^*\tilde{\mathcal{E}}) \right]\,,  \nonumber
\label{eq:sigmaX-TCS}
\end{eqnarray}
where $d\Omega_{l^+l^-} = d(\cos\theta) ~d\phi$.

The contribution of the interference between the TCS and BH mechanisms 
to the differential cross section for the $\gamma p \to l^+ l^- p$  process reads:
\begin{eqnarray}
\frac{d\sigma^{(INT)}}{dQ^2 dt d\Omega_{l^+l^-} d \Phi_h} 
& \, \equiv\,  & 
\frac{d\sigma^{(INT)}_{unpol}}{dQ^2 d...}+
\frac{d\sigma^{(INT)}_{linpol}}{dQ^2 d...} \,,
\end{eqnarray}
where:
\begin{eqnarray}
\frac{d\sigma^{(INT)}_{unpol}}{dQ^2 d...}
& \, =\,  &
\frac{\alpha^3}{16\pi^2s^2}\frac{1}{Q^2} \left(\frac{4s\mid{\Delta}_\perp\mid}{Qt}\right)\,
 \frac{1 + \cos^2\theta}{\sin\theta}\cos\phi 
{Re}\left[\mathcal{H}F_1 - \frac{t}{4M^2} \mathcal{E}F_2 - \eta\tilde{\mathcal{H}}(F_1 + F_2)) \right]  \,,\nonumber\\
\frac{d\sigma^{(INT)}_{linpol}}{dQ^2 d...} 
& \, =\,  &   
\frac{- \alpha^3}{16\pi^2s^2}\frac{1}{Q^2} \left(\frac{4s\mid{\Delta}_\perp\mid}{Qt}\right) \,  
          \sin\theta\cos(2\Phi_h + 3\phi) 
{Re}\left[\mathcal{H}F_1 - \frac{t}{4M^2} \mathcal{E}F_2 + \eta\tilde{\mathcal{H}}(F_1 + F_2)) \right] \,. \nonumber
\label{eq:sigmaX-INT}
\end{eqnarray}
%
%
The extraction of information from the interference part through the study of 
the angular distribution of the lepton pair relies on the following observation: if one reverses the lepton charge, which is equivalent to an exchange $k \leftrightarrow k^\prime$, the TCS and the BH amplitude transform with opposite signs. 
Thus, the cross sections associated with the TCS and the BH mechanism transform even under this exchange, but the interference cross section is odd. 
If one now considers observables which also change sign when exchanging $k$ and $k^\prime$, they give direct access to the interference term. 

We define two observables sensitive to the unpolarized and linearly polarized part of interference cross section. The first one is similar  to the $R$ ratio defined in \cite{BDP} in the case of an unpolarized photon beam :
\begin{equation}
\tilde{R} = 
\frac{\int_0^{2\pi} d\Phi_h 2\int_0^{2\pi} d\phi\cos(\phi) \int_{\pi/4}^{3\pi/4}\sin \theta d \theta\frac{d \sigma}{dt dQ^2 d\Omega d\Phi_h}}
{\int_0^{2\pi} d\Phi_h \int_0^{2\pi} d\phi \int_{\pi/4}^{3\pi/4}\sin \theta d \theta\frac{d \sigma}{dt dQ^2 d\Omega d\Phi_h}}\,.
\end{equation}
The second observable projects out the $d\sigma_{linpol}^{(INT)}$ part of the interference cross section:
\begin{equation}
\tilde{R}_3 = 
\frac{2\int_0^{2\pi} d\Phi_h \cos(2\Phi_h) 2\int_0^{2\pi} d\phi\cos(3\phi) \int_{\pi/4}^{3\pi/4}\sin \theta d \theta\frac{d \sigma}{dt dQ^2 d\Omega d\Phi_h}}
{\int_0^{2\pi} d\Phi_h \int_0^{2\pi} d\phi \int_{\pi/4}^{3\pi/4}\sin \theta d \theta\frac{d \sigma}{dt dQ^2 d\Omega d\Phi_h}}\,.
\end{equation}
Making use of $\tilde{R}$ and $\tilde{R}_3$ we can define the following observable which is sensitive only to the interference term and which provides us with information on $\tilde{\mathcal{H}}$:
\begin{equation}
C =  \frac{\tilde{R}}{\tilde{R}_3} =
\frac{2-3\pi}{2+\pi}
\frac
{{Re}\left[\mathcal{H}F_1 - \frac{t}{4M^2} \mathcal{E}F_2 - 	\eta\tilde{\mathcal{H}}(F_1 + F_2)\right]
}
{{Re}\left[\mathcal{H}F_1 - \frac{t}{4M^2} \mathcal{E}F_2 + 	\eta\tilde{\mathcal{H}}(F_1 + F_2)\right]
}\,.
\label{Cdef}
\end{equation}

\begin{figure}
\includegraphics[keepaspectratio,width=0.45\textwidth,angle=0]{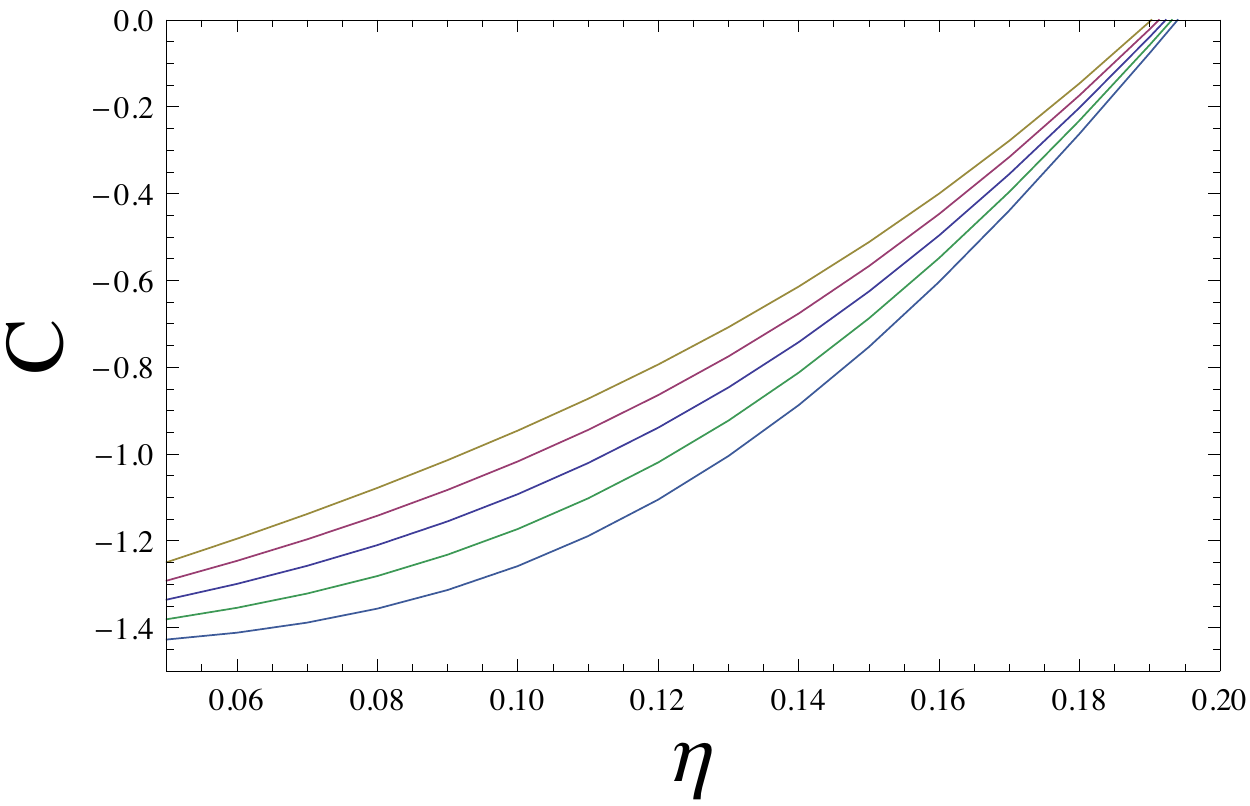} 
\hspace{0.05\textwidth}
\includegraphics[keepaspectratio,width=0.45\textwidth,angle=0]{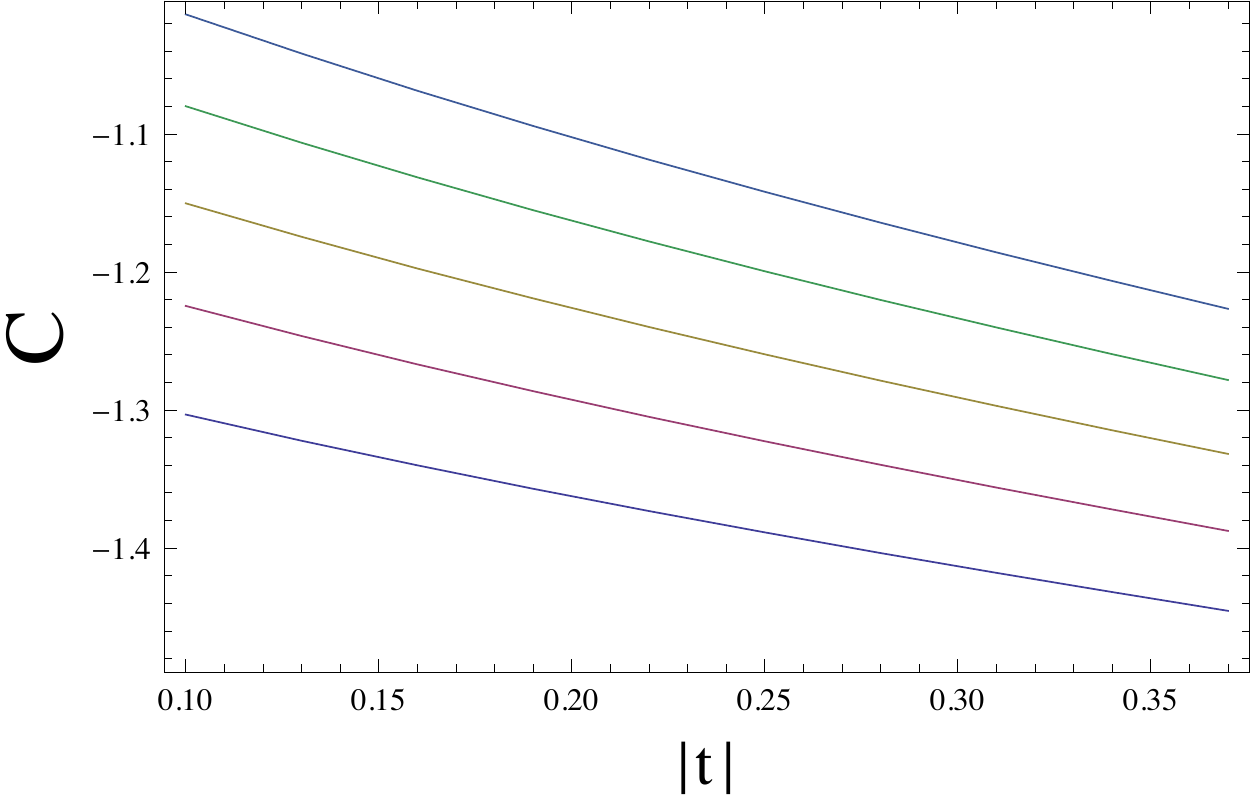}
\caption{(Left) $C$ as a function of $\eta$ calculated for $Q^2=4 \gev^2$ and $t= t_0(\eta)$. (Right) $C$ as a function of $t$ for  $Q^2=4 \gev^2$ and $\eta= 0.1$. Different curves correspond to different polarized GPD models $\tilde{H}_g = \{-1,0,1,2,3\} \cdot \tilde{H}_g^{GK}$. Calculations where performed with NLO accuracy and $\alpha_S=0.3$.}
\label{fig:tildeC}
\end{figure}

In order to illustrate the dependence of the proposed observable on $\tilde{H}$ we make use of the GK GPDs model based on fits to Deeply Virtual Meson Production \cite{GK}.  We present our numerical estimates with the GPD $E$ set to zero, as it is largely unknown and also suppressed by the small kinematic factor $\frac{t}{4M^2}$. 
On the left-hand side of figure \ref{fig:tildeC} we present the ratio $C$ as a function of $\eta$, calculated for $Q^2 = 4 \gev^2$ and a minimal value of $t= t_0(\eta) = -4\eta^2 M^2/(1-\eta^2)$. To study the sensitivity of $C$ on the GPD $\tilde{H}$ we have varied the gluonic contribution, taking it as $\tilde{H}_g = \{-1,0,1,2,3\}\cdot \tilde{H_g}^{GK}$, where $\tilde{H_g}^{GK}$ is given by the GK model. On the right-hand side of figure \ref{fig:tildeC} we show $C$ as a function of $t$ for $Q^2 = 4 \gev^2$ and $\eta =0.1$. In both cases significant effects of the order of $30\%$ are visible proving that $C$ is a good observable sensitive to $\tilde{H}$.  
\section{Electroproduction with a small electron scattering angle.}
In the unpolarized electron scattering process, the virtual photon polarization is given by \cite{Dombey:1969wk}:
\begin{equation}
\epsilon = \left[1+ 2\frac{Q_\gamma^2+\nu^2}{Q_\gamma^2}\tan^2 (\theta_{e'}/2)\right]^{-1} \nonumber
\end{equation}
where $\nu$ is the photon energy and $\theta_{e'}$ the electron scattering angle. The longitudinal polarization
 is given by $\epsilon_L = \frac{Q_\gamma^2}{\nu^2}\epsilon$ and in the limit $Q_\gamma^2<<\nu^2 $ the polarization density matrix
\begin{equation}
\rho = \left(
\begin{array}{ccc}
\frac{1}{2}(1+\epsilon)
&
0
&
\sim \epsilon_L^{1/2}
\\
0&\frac{1}{2}(1-\epsilon)&0 \\
\sim \epsilon_L^{1/2}&0&\epsilon_L
\end{array}
\right)\nonumber	\qquad,
\end{equation}
describes real transverse photons. The cross section 
\begin{equation}
\frac{d \sigma^{ep}}{d Q^2_\gamma d \nu} \sim F(Q_\gamma^2,\nu)\, \mathcal{M}_\lambda^* \,\, \rho^{\lambda \lambda '} \mathcal{M}_{\lambda'}\, \nonumber
\end{equation}
is given by the incoherent sum of two polarizations:
\begin{eqnarray}
\frac{d \sigma^{ep}}{d Q^2_\gamma d \nu} &=& F(Q_\gamma^2,\nu)\,\, 
\left[
 \frac{1}{2}(1+\epsilon) \sigma^{xx}_{\gamma p}
+\frac{1}{2}(1-\epsilon) \sigma^{yy}_{\gamma p}
\right] \nonumber \\&=&
F(Q_\gamma^2,\nu)\,\, 
\left[
(1-\epsilon) \sigma^{unp}_{\gamma p}
+\epsilon \, \sigma^{linpol}_{\gamma p}
\right]\nonumber
\end{eqnarray}
where $F(Q_\gamma^2,\nu)$ is the photon flux. In that case we can define the observable, analogous to the one defined by the Eq.(\ref{Cdef}), in the following way:
\begin{equation}
C^{ep} = \frac{\tilde{R}^{ep}}{\tilde{R}^{ep}_3} = \frac{1-\epsilon}{\epsilon} \,C \,. \nonumber
\end{equation}

\section*{Acknowledgements} 
This work is partly supported by the Polish NCN Grant No. DEC-2011/01/D/ST2/02069 and the Austrian FWF Grant No. J 3163-N16.


\end{document}